\documentclass{article}
\usepackage{spconf,amsmath,epsfig}

\let\OLDthebibliography\thebibliography
\renewcommand\thebibliography[1]{
  \OLDthebibliography{#1}
  \setlength{\parskip}{0pt}
  \setlength{\itemsep}{0pt plus 0.3ex}
}

\usepackage{subfigure}

\begin{document}\sloppy

\def\x{{\mathbf x}}
\def\L{{\cal L}}

\title{Depression Diagnosis and Analysis via Multimodal Multi-order Factor Fusion}
%
\name{Chengbo Yuan$^{\dagger}$, Qianhui Xu$^{\dagger}$ and Yong Luo$^{\dagger}$}
\address{}
\address{$^{\dagger}$Wuhan University, \{michael.yuan.cb,xuqianhui,luoyong\}@whu.edu.cn}

\maketitle

\begin{abstract}
Depression is a leading cause of death worldwide, and the diagnosis of depression is nontrivial.
Multimodal learning is a popular solution for automatic diagnosis of depression, and the existing works suffer two main drawbacks: 1) the high-order interactions between different modalities can not be well exploited; and 2) interpretability of the models are weak. To remedy these drawbacks, we propose a multimodal multi-order factor fusion (MMFF) method. Our method can well exploit the high-order interactions between different modalities by extracting and assembling multi-order factors across modalities under the guide of a shared latent proxy.
We conduct extensive experiments on two recent and popular datasets, E-DAIC-WOZ and CMDC, and the results show that our method achieve significantly better performance compared with other existing approaches. Besides, by analyzing the process of factor assembly, our model can intuitively show the contribution of each factor. This helps us understand the fusion mechanism. 
\end{abstract}
\begin{figure*}[!t]
	\centering
	\includegraphics[scale=0.125]{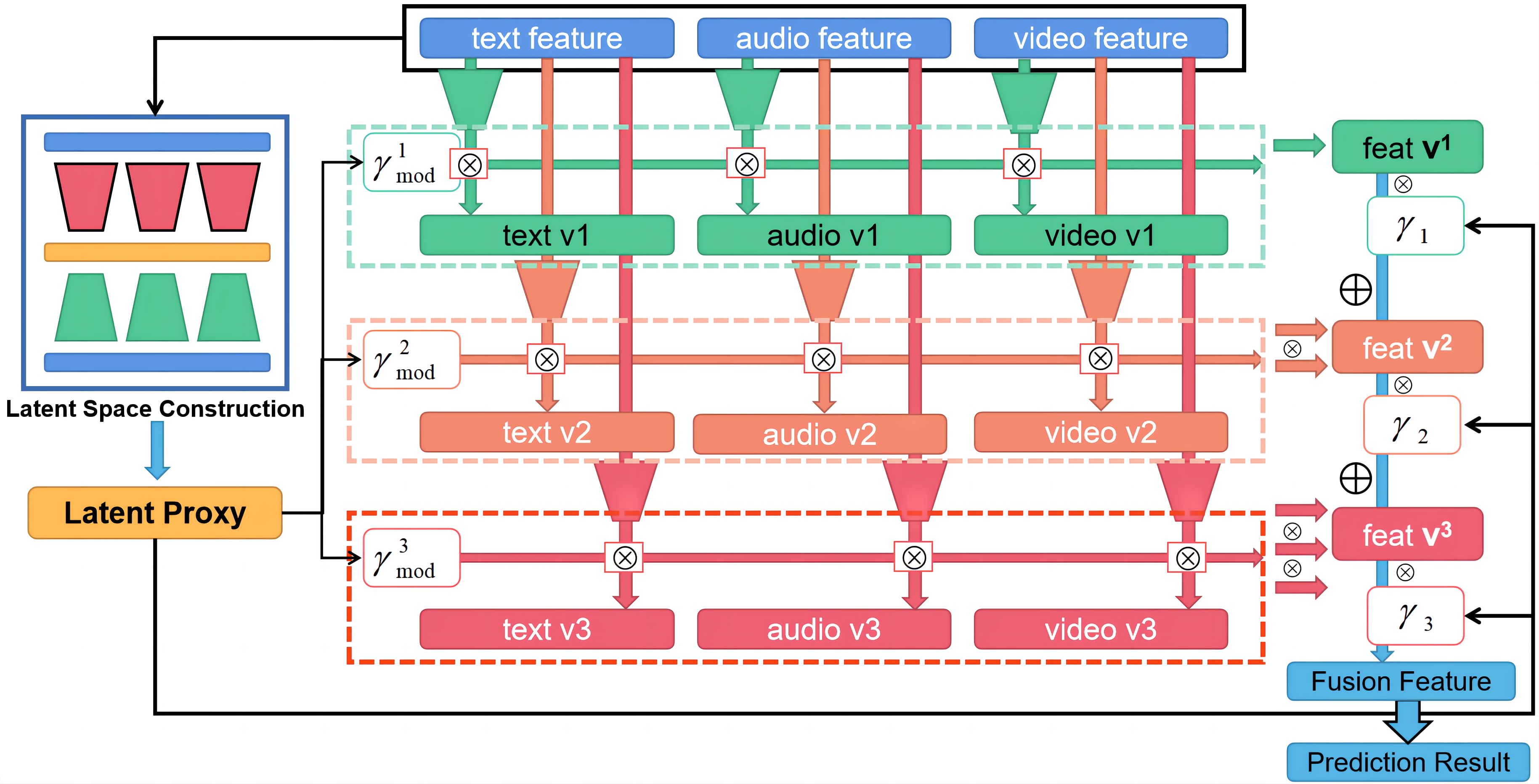}
	\caption{Illustration of the proposed Multimodal Multi-order Factor Fusion (MMFF) method. Specifically, our MMFF contains two main parts: 1) a latent space construction module based on the encoder-decoder mechanism to extract a proxy, which is utilized to generate adaptive weights to fuse different modality factors; 2) a multi-order combination part that applies multiple encoders to generate factors, and integrate them under the guidance of the shared-subspace proxy at different levels of orders.}
	\label{fig:factor}
\end{figure*}

\section{Introduction}
\label{sec:intro}
Depression is a common and serious mental disorder disease around the world. It is estimated that about $3.8\%$ of the population is affected by the disease, and more than $700,000$ patients die of suicide every year due to depression, according to the statistics of world health organization (WHO). Due to the exhaustive consumption of human resources and heavy dependence of subject judgement of the traditional diagnosis approaches, there is an increasing interests in utilizing machine learning for automatic diagnosis of depression, and multimodal learning is one of the most popular solutions.

The existing multimodal learning works either focus on improving the feature extraction for different modalities, or designing better fusion strategies.
To extract features for long-time sequential data,
some classical models such as LSTM and CNN are adopted in~\cite{AVEC2019,BERTCNN,MS}, while more recent approaches employ transformer to alleviate the forgeting problem caused by the long duration of patient interview~\cite{ATransformer}.
In regard to the modality fusion, in addition to the simple concatenation, attention mechanism is applied to adjust the modality contribution~\cite{DepressNet}. Some other approaches, such as CubeMLP~\cite{CubeMLP} and Multi-Agent~\cite{MultiAg} can exploit more complex interactive relationships between different modalities.
Although have achieved certain success, these approaches suffer from two main drawbacks: 1) lack of exploitation of different-order interactions across modalities;
and 2) the realtively low interpretability, which leads to the confusion of the fusion mechanism.

To remedy these drawbacks, we propose a Multimodal Multi-order Factor Fusion method (MMFF) based on modality multi-order factor extraction and assembly. The idea of MMFF is to first extract multi-order factors across different modalities and then integrate the factors in terms of different orders for final prediction.
The whole process is under the guidance of a shared subspace proxy, which is utilized to generate adaptive weights for modality factor fusion at different orders. The learned weights can intuitively show the contribution of each factor in the fusion process.
Besides, we extract features for the long time sequential data (such as videos and audios) from a new perspective based on trajectory preservation, where a novel frame selection pipeline based on the principal component analysis (PCA) and Midimax algorithm is developed. This helps us to effectively select the key frames and compress the time series.


To summarize, our main contributions are as follows:
\begin{itemize}
\item We propose a novel multimodal multi-order factor fusion (MMFF) method for the diagnosis of depression. The proposed method is able to exploit the modality interactions at different levels of orders.
\item We design a latent proxy to generate adaptive weights for different modality factors (and also different orders). This improves the interpretability of our model since the contribution of each factor can be identified at different levels of orders.
This enables the fusion mechanism analysis.
\item We develop a novel method to select key frames for long time sequential data by integrating the PCA and Midimax algorithm.
\end{itemize}
The experiments are conducted on two recent datasets: the Extended Distress Analysis Interview Corpus (E-DAIC-WOZ)~\cite{AVEC2019,DAIC} and the Chinese Multimodal Compression Corpus (CMDC)~\cite{CMDC}.
The results demonstrate that our MMFF is superior to the state-of-the-art approaches with significant improvements.

\section{The Proposed Method}
Figure~\ref{fig:factor} is an illustration of the overall structure of the proposed Multimodal Multi-order Factor Fusion (MMFF) method. In this section, we first introduce the feature extractors for different modalities, where a novel frame selection pipeline is developed.
Then an encoder-decoder module is utilized to construct a proxy to generate adaptive weights for combination.
Finally, the factors extracted from different modalities are integrated at different levels of orders, and the integrated representations are further adaptively fused for final prediction.

\subsection{Feature Extraction} 
In this paper, we use a classic sentence to vector (sent2vec) encoder, the Universal Sentence Encoder~\cite{USE} together with BiLSTM at the sentence level~\cite{AU2} to extract the textual features for the patient's interview.
We use the pretrained USE to extract the sentence embedding, and then apply a two-layer BiLSTM and three-layer MLP to extract the features.

\begin{figure}[!t]
	\centering
	\includegraphics[scale=0.13]{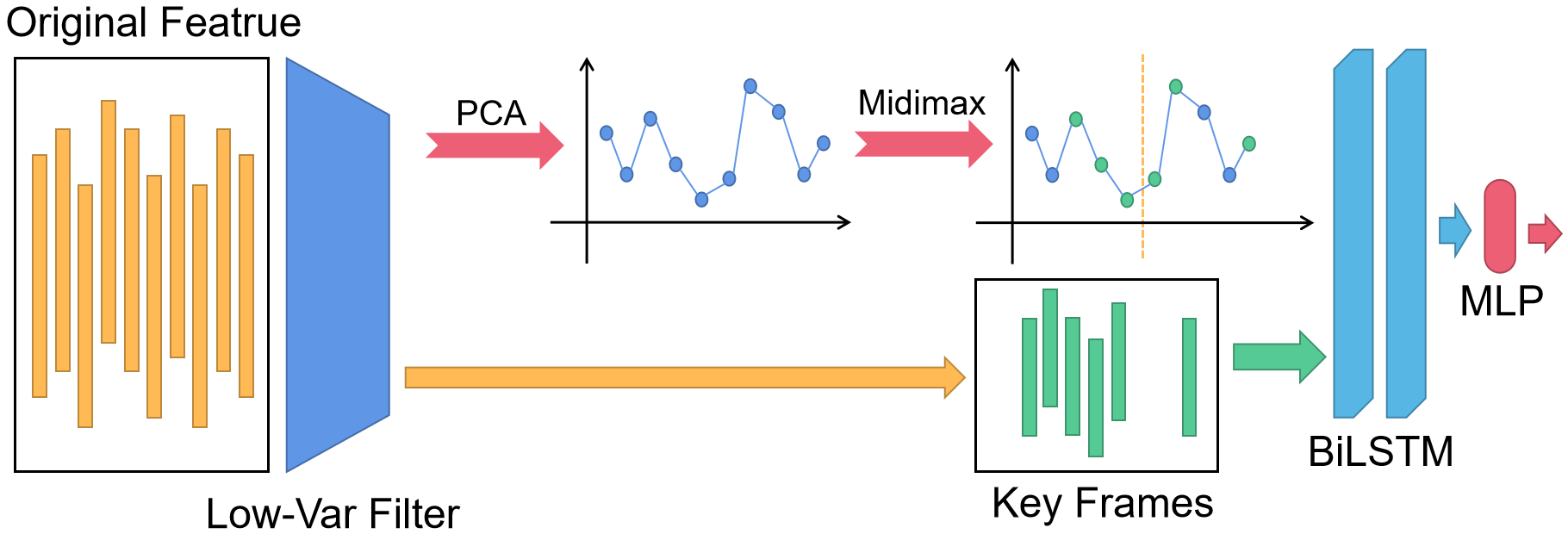}
	\caption{The proposed frames feature extractor. Low variance filter is firstly used to remove the insignificant feature. Then, the Midimax algorithm is applied to the first principle component of the original feature to acquire the key frames.}
	\label{fig:frame}
\end{figure}

In regard to the audio and video inputs, we utilize the low-variance filtering, PCA and Midimax algorithms together with a two-layer BiLSTM to extract the features.
The process is illustrated in Figure~\ref{fig:frame}, where the original features for each video frame are extracted following~\cite{AVEC2019,DAIC} for the E-DAIC-WOZ dataset.
For the CMDC dataset, the original feature is a stacking of the $12$ feature vectors, each corresponds to a certain question in the interview.
Since the number of frames in each video or audio sequence may be very large (such as in the E-DAIC-WOZ dataset), the computational cost may be very high and there may be much redundant information. Therefore, we process these original features using the following strategy.

Firstly, low-variance filtering is performed on the original feature. Suppose that the original feature after min-max normalization is $\{X^{K \times M_i}_i=[X_i^1,X_i^2,\cdots,X_i^{M_i}],\ i=1,2...N\}$, where $N$ is the size of training set, $X_i^j$ is the $j$-th frame of $i$-th sample, $K$ is the feature dimension of the frame, $M_i$ is the number of frames in $X_i$, and let $X_i^{j}=[x_i^{j1}, x_i^{j2}, \cdots, x_i^{jK}]^T$, then the $k$-th dimension of features that satisfies the following condition is filtered:
\begin{equation}
\frac{1}{N}\sum_{i=1}^N \frac{1}{M_i}\sum_{j=1}^{M_i} (x_i^{jk}-E_j[x_i^{jk}])^2 \leq \beta,
\end{equation}
where $\beta$ is an hyperparameter (we set $\beta=0.01$ in this paper).
Specifically,  $\frac{1}{M_i}\sum_{j=1}^{M_i} \left(x_i^{jk}-E_j[x_i^{jk}]\right)^2$ is the variance of $X_i$ at the $k^{th}$ dimension.
After the low variance filtering, some redundant features are removed.

Then we compress the sequence based on variance extraction and the Midimax algorithm. The main idea of Midimax algorithm is to keep the trajectory of sequence as much as possible under the premise of satisfying a certain compression ratio. Considering that Midimax is mainly applicable to one-dimensional data, we first perform the maximum variance projection of features. That is, we use the PCA algorithm to extract one principal component to transform the sequence to one-dimension, and then apply Midimax to the projected sequence.

\begin{figure}[!t]
	\centering
	\includegraphics[scale=0.15]{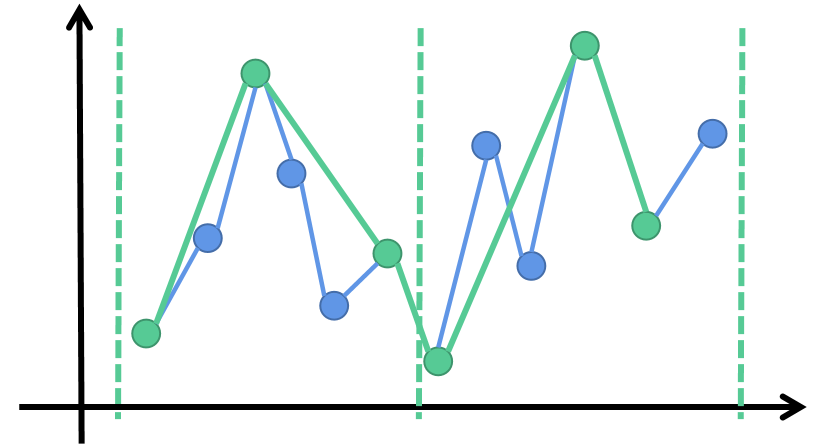}
	\caption{The Midimax Algorithm ($\delta$=6). To preserve the original trajectory, the maximum, minimum and median are picked up for each time slice under the given compression ratio.}
	\label{fig:midimax}
\end{figure}

The following is a brief description of the Midimax algorithm we apply: given the compression ratio $\delta$, which is an integer, we first divide the sequence into $[s_1,s_2...s_{Ns}]$, where $Ns$ is the number of time slices, $s_i$ is the $i^{th}$ slice, and the length of each time slice is $\delta$. Let $s_i=[s_i^1,s_i^2...s_i^{\delta}]$, and suppose that the index of the maximum, minimum and median in the time slice are $p_1$, $p_2$, $p_3$. Then we sort $[p_1,p_2,p_3]$ as $[q_1,q_2,q_3]$ ($q_1\leq q_2\leq q_3$) and finally compress $s_i$ into $[s_i^{q_1},s_i^{q_2},s_i^{q3}]$. The scheme is shown in Figure \ref{fig:midimax}. Finally, BiLSTM is applied to extract the features.

\subsection{Proxy Construction}
\begin{figure}[!t]
	\centering
	\includegraphics[scale=0.16]{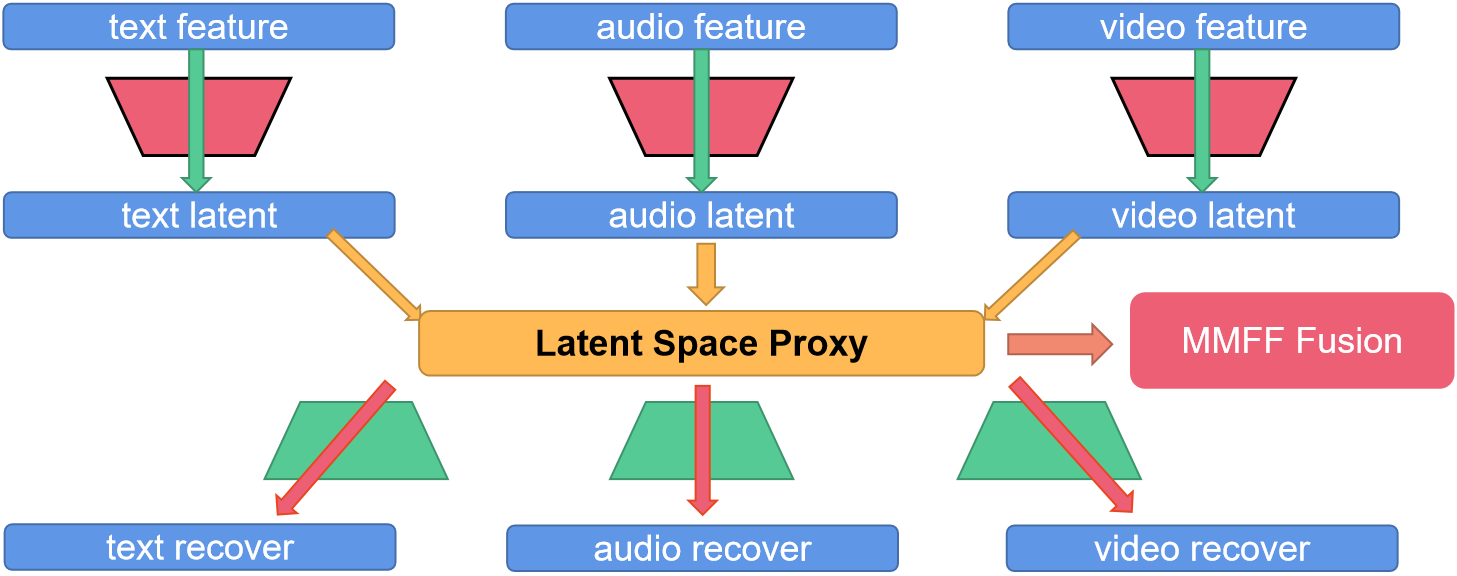}
	\caption{Illustration of the latent space module for proxy extractor. The encoder-decoder form is adopted to maximize the sharing between modality subspace while preserving the modality information.}
	\label{fig:latent}
\end{figure}

In factor fusion, we need to calculate the factor contributions for each modality.
Considering that the calculation of factor contribution should be consistent, we first map three modalities into a common low-dimensional subspace, which is learned by reconstructing the inputs with minimum errors.
That is, in order to make the new subspace effectively represent the original space, a reverse projection is conducted to map the subspace back to the original space, and employ the mean square error (MSE) of the recovery space and the original space as the loss function, as illustrated in Figure~\ref{fig:latent}.

Formally, let $x_{t}$, $x_{a}$, $x_{v}$ be the extracted features for the textual, audio and visual modalities respectively, we learn a mapping function $F(\cdot)$, and the subspace mapping can be expressed as:
\begin{equation}
    \begin{split}
	& y_{mod} = F_{mod}(x_{mod}),\ \ mod=\{t,a,v\},\\
	& z=\frac{1}{3}(y_{t} + y_{a} + y_{v}),
	\end{split}
\end{equation}
where the dimension of $y_{mod}$ and the latent representation $z$ is less than $x_{mod}$. The latent subspace is learned by using the following loss function:
\begin{equation}
    \begin{split}
	& \mathcal L_{latent} = \frac{1}{3} \sum_{mod\in \{t,a,v\}}(\hat{x}_{mod} - x_{mod})^2,\\
	& \hat{x}_{mod} = F'_{mod}(z).
	\end{split}
\end{equation}
The learned $z$ is utilized as a proxy to guide the subsequent factor fusion.

\subsection{Factor Extraction \& Assembly}
There are many approaches for multimodal fusion, but most of them have two main issues: the incomplete exploitation of the interactive information across modalities and low interpretability. Our MMFF model alleviates these issues by exploiting multi-order interactions between different modalities in the combination, which also provides a certain degree of interpretability for fusion mechanism analysis.

\begin{figure}[!t]
	\centering
	\includegraphics[scale=0.2]{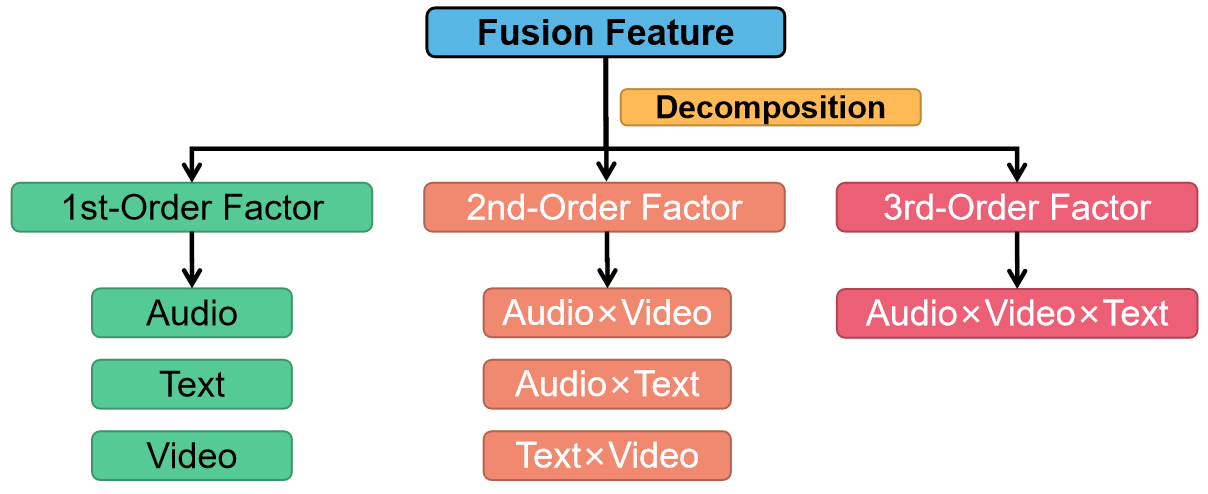}
	\caption{The proposed contribution decomposition methodology. The core idea is to decompose the interaction gain to all modality combination explicitly, so that the fully coverage of interaction extraction is assured.}
	\label{fig:decomp}
\end{figure} 
		
The factor fusion process at different levels of orders is illustrated in Figure~\ref{fig:decomp}. First, the features of each modality are decomposed into factors of different orders.
where the $p$-order factor indicates that the factor comes from $p$ modalities.
Then for the $p$-order factor, there will be $C_{3}^{p}$ subfactors, where $C_{3}^p$ is the combination number. How to extract each factor will be depicted as follows.
		
Let $mod$ denote a single modality, $com$ signifies modality combination, and $H(\cdot)$ be a certain mapping function, we first calculate the contribution of each modality for the factors of different levels of orders.
This is achieved by using the constructed proxy to generate the adaptive weights.
Specifically:
\begin{equation}
	\gamma^k = H_k(z),\ \ k=1,2,3,
\end{equation}
where $\gamma^k=[\gamma_t^k,\gamma_{a}^k,\gamma_{v}^k]^T$, which satisfies $\gamma^k_t+\gamma^k_a+\gamma_v^k=1$. The 1st-order factor $v_{mod}^1,\ mod\in \{t,a,v\}$, is calculated as follows:
\begin{equation}
    v_{mod}^1 = \gamma^{1}_{mod}P_{mod}^T G_1^{mod}(x_{mod}),\ \ mod=\{t,a,v\},
\end{equation}
where $P_{mod}^T$ is a projection matrix, $G(\cdot)$ is an encoder and $\gamma^{1}_{mod}$ is the contribution of certain modality at the 1st-order.
Given the contribution $\gamma_{mod}^{2}$ for each modality at the 2nd-order, we use the multimodal low-rank bilinear (MLB) approach~\cite{mlb} to construct the 2nd-order factor, i.e.,
\begin{equation}
    \begin{split}
    & v_{ta}^2 = P_{ta}^T\sigma(\gamma_{t}^2 G_{2}^t(x_t) \circ \gamma_{a}^2 G_{2}^a(x_a)),\\
    & v_{av}^2 = P_{av}^T\sigma(\gamma_{a}^2G_{2}^a(x_a) \circ \gamma_{v}^2G_{2}^v(x_v)),\\
    & v_{tv}^2 = P_{tv}^T\sigma(\gamma_{t}^2G_{2}^t(x_t) \circ \gamma_{v}^2G_{2}^v(x_v))
	\end{split}
\end{equation}
where $\sigma(\cdot)$ is a certain activation function and $\circ$ is the Hadamard product. Then the MLB approach is extended to induce the 3rd-order factor:
\begin{equation}
	v_{tav}^3 = P_{tav}^T \sigma(\gamma_{t}^3G_{3}^t(x_t) \circ \gamma_{a}^3G_{3}^a(x_a) \circ \gamma_{v}^3G_{3}^v(x_v)).
\end{equation}
We then combine the obtained subfactors at different levels of orders. The integrated 1st-order factor $v^1$, 2nd-order factor $v^2$ and 3rd-order factor $v^3$ are obtained by simply adding the subfactors and then performing a projection, i.e.,
\begin{equation}
    \begin{split}
	& v^1 = P_1^T(v^1_{t} + v_a^1 + v_v^1),\\
    & v^2 = P_2^T(v^2_{ta} + v_{av}^2 + v_{tv}^2),\\
    & v^3 = P_3^T v_{tav}^3.
	\end{split}
\end{equation}

The final fusion feature is composed of all the 1st, 2nd and 3rd-order factors. Similarly, we use the proxy to calculate the contribution of each order, and then weightedly fuse the integrated factors to obtain the final feature $v$:
\begin{equation}
    v = \gamma_1v^1 + \gamma_2v^2 + \gamma_3v^3.
\end{equation}
Here,
\begin{equation}
	\gamma = [\gamma_1,\gamma_2,\gamma_3]^T= H_{com}(z),\ \ s.t.\ \ \sum_{i=1}^3 \gamma_i = 1.
\end{equation}

For our MMFF, A hierarchical training strategy is adopted. That is, we first train the latent subspace solver, and then freeze its parameters, and finally train the backbone of our MMFF. After obtaining the final fusion feature $v$, we use a three-layers MLP to predict patients' PHQ score. Let $\hat{y}$ be the prediction score, $y$ be the real label, and the loss function used in our training process is the MSE loss function, i.e.,
\begin{equation}
	\mathcal L_{MSE} = \frac{1}{N} \sum_{i=1}^N (\hat{y} - y)^2.
\end{equation}
Through the adaptive training of the contributions $\gamma^1$, $\gamma^2$, $\gamma^3$ and $\gamma$, we can intuitively acquire the importance of each factor of modality in the combination, so as to better understand the mechanism of fusion.

\section{Experiments}

\subsection{Dataset}
Our experiments are mainly conducted on two recent depression diagnosis datasets, namely, the Extended Distress Analysis Interview Corpus (E-DAIC-WOZ)~\cite{AVEC2019,DAIC} and Chinese Multimodal Compression Corpus (CMDC)~\cite{CMDC}.

E-DAIC-WOZ is an extended version of WOZ-DAIC dataset\cite{DAIC}, and it is also the dataset for the AVEC2019 DDS Challenge~\cite{AVEC2019}.
E-DAIC-WOZ provides video, audio and text modality data. For the text modality, the interview records are provided. For the video modality, only the extracted features are provided to protect the privacy of patients. For audio modality, in addition to the wav file, some features are also provided.
The dataset contains $163$ training samples, $56$ validation samples and $56$ test samples.

CMDC is a newly proposed dataset, which is collected by \cite{CMDC}.
This dataset was obtained from the audio-visual records of patients' clinical interviews in the Chinese environment.
CMDC presets 12 questions during the interview and stores the answering data for these 12 questions.
Similarly, in order to ensure the privacy of patients, the datasets only provided extracted features in the audio and video modalities.
The dataset consists of $45$ samples that have complete modalities: $19$ samples for patients with depression and $26$ samples for the control group. Considering the relatively small amount of data, the author provides a stratified 5-fold cross validation to obtain the final performance of the model.

\subsection{Experiment Setup}
Our experimental settings are as follows: the activation function for both the audio and video modalities are ELU($\cdot$), and the activation function for the text modality is ReLU($\cdot$). In the factor fusion part, $F(\cdot)$ is a two-layer MLP with Tanh activation, $G(\cdot)$ is a single-layer MLP with Hardtanh activation, and $H(\cdot)$ is a three-layer MLP with ReLU activation.
In order to alleviate the over-fitting problem, we use the AdamW optimizer for training, and add the Dropout layer to the network backbone. The dropout rate is set between $[0.4,0.6]$.

For the E-DAIC-WOZ dataset~\cite{AVEC2019,DAIC}, we follow the data partition in AVEC2019 DDS. We use Midimax to compress the length of video and audio feature to $1200$ and $600$ respectively, and unit normalization is conducted on the compressed results.
For the CMDC dataset \cite{CMDC}, we first remove the data with incomplete modalities as described in Zou et al.\cite{CMDC}, and then divide the remaining data into five folds according to the stratified partition provided in the CMDC attached file. Then, for the non-test data in each fold, we randomly take $80\%$ data as the training set, and set the remaining data as the validation set. The model is evaluated based on the combination of prediction scores from five folders.

In the AVEC2019 DDS~\cite{AVEC2019}, CCC (concordance correlation coefficient) and RMSE (the root mean square error) are the main evaluation metrics. To test the effectiveness of our model, we add the MAE (mean absolute error) as an extra criterion.
For the CMDC dataset, we follow~\cite{CMDC} to adopt the RMSE, MAE and Pearson correlation coefficients as the evaluation criteria.


\subsection{Experimental Results}
The results on the E-DAIC-WOZ~\cite{AVEC2019,DAIC} dataset are reported in Table \ref{tab:DAIC}. From the results, we observe that: 1) after low-variance filtering and Midimax processing, our model obtains CCC=$0.434$ and RMSE=$5.44$ using only the audio modality. This is a competitive result, which verifies the effectiveness of our Midimax sequence compression; 2) the proposed MMFF achieves the performance of CCC=$0.676$, RMSE=$4.91$, and MAE=$3.98$. It can be seen that our model achieves the best performance on the most popular evaluation metric CCC, and ranked first and third in terms of $MAE$ and $RMSE$ respectively. This demonstrates the superiority of our method.

\begin{table}[!t]
   	\centering
   	\caption{Result on E-DAIC-WOZ Test Set, where the best results are marked in bold, ``(T)'' and ``(A)'' indicate that the model only uses the corresponding modality. If not specified, all modalities are utilized.
   	} 
   	\label{tab:DAIC}
   	\resizebox{.9\columnwidth}{!}{
   	\begin{tabular}{|l|l|l|l|}
   	\hline
   	Model                       & \textbf{CCC}   & RMSE & MAE  \\ \hline
   	Baseline~\cite{AVEC2019}                   & 0.111 & 6.37 & /    \\ 
   	Adaptive Transformer~\cite{ATransformer} & 0.331 & /    & 6.22 \\ 
   	Bert-CNN \& Gated-CNN~\cite{BERTCNN}       & 0.403 & 6.11 & /    \\ 
   	VarFilter+Midimax (Ours) (A)          & 0.434 & 5.44 & 4.4 \\ 
   	Hierarchical BiLSTM~\cite{HR}         & 0.442 & 5.50 & /    \\ 
   	DepressNet~\cite{DepressNet}                  & 0.457 & 5.36 & /    \\ 
   	multi-DAAE+Fusion~\cite{AU2}   & 0.528 & \textbf{4.47} & /    \\ 
   	PV-DM~\cite{AU2} (T)   & 0.560 & 4.66 & /    \\ 
   	CubeMLP~\cite{CubeMLP}                 & 0.583 & /    & 4.37 \\ 
   	MMFF (Ours)          & \textbf{0.676} & 4.91 & \textbf{3.98} \\ \hline
   	\end{tabular}
   	}
\end{table}
   	
The results on the CMDC dataset are reported in Table \ref{tab:CMDC}. In the experiments of Zou et al.~\cite{CMDC}, the combination including the audio modality achieves better performance compared with others. This is in line with the results in our experiments. The performance of audio modality is surprisingly good. By applying low variance filtering, normalization and 2-layer BiLSTM, utilizing only the audio modality can outperform~\cite{CMDC}.
By combining the different modalities using our MMFF method, the performance can be further improved.

\begin{table}[!t]
   	\centering
   	\caption{Result on CMDC Test Set, where the best results are marked in bold; ``(A)'' and ``(A+T)'' indicate that the model only uses the corresponding modalities. If not specified, all modalities are utilized.} 
   	\label{tab:CMDC}
   	\resizebox{.9\columnwidth}{!}{
   	\begin{tabular}{|l|l|l|l|}
   	   	\hline
   	   	Model                     & Perason & RMSE & MAE  \\ \hline
   	   	SVR                       & 0.57    & 7.04 & 5.31 \\ 
   	   	KNN                       & 0.50    & 8.38 & 5.89 \\ 
   	   	BiLSTM~\cite{CMDC}      & 0.68    & 5.67 & 4.55 \\ 
   	   	MulT~\cite{CMDC}  & 0.72    & 5.61 & 4.32 \\ 
   	   	MulT~\cite{CMDC} (A+T)          & 0.72    & 4.59 & 3.66 \\ 
   	   	VarFilter+2BiLSTM (Ours) (A) & 0.81    & 4.31 & 3.22 \\ 
   	    MMFF (Ours)        & \textbf{0.83}    & \textbf{4.29} & \textbf{3.19} \\ \hline
   	\end{tabular}
   	}
\end{table}
   	
\subsection{Ablation Study for Factor Order}

In order to further evaluate our model and understand the fusion mechanism, we conducted ablation studies on the E-DAIC-WOZ dataset to verify the role of each order in factor fusion.
The results are shown in Table \ref{tab:aborder}, we can be see that: 1) the full set of factor combinations performs the best in terms of CCC, the combination of 1st-order and 3rd-order factors achieves the best performance on RMSE, and the utilization of only the 1st-order factors performs the best on MAE. In general, involving more levels of order combination tend to improve the performance, and this demonstrates the necessity of integrating factors of different orders; 2) the CCC of all ablation models are higher than $0.65$, which demonstrates the reliability of our model.


\begin{table}[!t]
   	\centering
   	\caption{Result of Factor Order Ablation Study, where the best results are marked in bold.}
   	\label{tab:aborder}
   	\resizebox{.9\columnwidth}{!}{
   	\begin{tabular}{|l|l|l|l|}
   		   	\hline
   		   	Factor Order Combination    & CCC   & RMSE & MAE  \\ \hline
   		   	1st + 2nd + 3rd & \textbf{0.676} & 4.91 & 3.98 \\ 
   		   	1st + 2nd       & 0.672 & 4.82 & 3.91 \\ 
   		   	1st + 3rd       & 0.674 & \textbf{4.80} & 3.92 \\
   		   	2st + 3rd       & 0.660 & 5.20 & 4.05 \\
   		   	1st             & 0.661 & 4.83 & \textbf{3.90} \\
   		   	2st             & 0.658 & 5.24 & 4.10 \\ 
   		   	3st             & 0.656 & 4.89 & 3.95 \\ \hline
   		   	\end{tabular}
   	}
\end{table}

\subsection{Fusion Mechanism Analysis}
As mentioned above, the contribution rate ($\gamma^1$,$\gamma^2$,$\gamma^3$,$\gamma$) obtained from the proxy can provide us with the mechanism explanation of factor fusion to some extent. In this section, we visualize the modality contributions on the E-DAIC-WOZ dataset and analyze the fusion mechanism.

The weights of each modality on each order and the whole process are shown in Table \ref{tab:contmodal}. It can be seen that the audio (42.5\%) is the most important modality, followed by the text (30.4\%) and video (27.1\%). We also observe that the audio, text, and video modalities have the largest contribution for the 1st-order, 2nd-order, and 3rd-order factors, respectively.
This indicates that the contributions of different modalities vary when exploiting the interactions in terms of different orders.

Besides, the contribution of audio modality at the 1st, 2nd and 3rd-order decreases ($56.6\%$ to $33.0\%$), and the contribution of text at the 2nd and 3rd-order decreases ($40.8\%$, $27.7\%$), while the contribution of video at the 1st, 2nd and 3rd-order increases ($18.9\%$ to $39.3\%$). Therefore, we speculate that there is a layer-by-layer extraction mechanism in our MMFF. That is, the information extracted by order $t$ is mainly the information that has not be fully extracted by the order smaller than $t$, thus realizing the hierarchical and full extraction of modality information. 

\begin{table}[!t]
   	\centering
   	\caption{Modality Contribution learned by our MMFF, where the largest contribution is marked in bold.}
   	\label{tab:contmodal}
   	\resizebox{.7\columnwidth}{!}{
   	\begin{tabular}{|l|l|l|l|}
   		   	\hline
   		   	Factor     & text   & audio  & video  \\ \hline
   		   	1st factor & 24.5\% & \textbf{56.6}\% & 18.9\% \\ \hline
   		   	2nd factor & \textbf{40.8}\% & 34.7\% & 24.5\% \\ \hline
   		   	3rd factor & 27.7\% & 33.0\% & \textbf{39.3}\% \\ \hline
   		   	fusion     & 30.4\% & \textbf{42.5\%} & 27.1\% \\ \hline
   		   	\end{tabular}
   	}
\end{table}

\section{Conclusion}
In this paper, we propose a Multimodal Multi-order Factor Fusion (MMFF) method for depression diagnosis. Comprared with the existing approaches, our method can ensure the full coverage of interactive information in different orders, while providing stronger interpretability.
Experiments conducted on two recent depression datasets show that our MMFF is significantly superior to the state-of-the-art approaches. We also conduct analysis of the learned contribution of each factor and give a conjecture of the fusion mechanism (namely, order-hierarchical information extraction for fusion).
In the future, we intend to apply the proposed method to more applications in addition to the depression diagnosis.

\bibliographystyle{IEEEbib}
\bibliography{icme2023_MMFF}

\begin{thebibliography}{10}

\bibitem{AVEC2019}
Fabien Ringeval, Bj{\"o}rn Schuller, Michel Valstar, Nicholas Cummins, Roddy
  Cowie, Leili Tavabi, et~al.,
\newblock ``Avec 2019 workshop and challenge: state-of-mind, detecting
  depression with ai, and cross-cultural affect recognition,''
\newblock in {\em AVEC2019}, pp. 3--12.

\bibitem{BERTCNN}
Mariana Rodrigues~Makiuchi, Tifani Warnita, Kuniaki Uto, and Koichi Shinoda,
\newblock ``Multimodal fusion of bert-cnn and gated cnn representations for
  depression detection,''
\newblock in {\em AVEC2019}, pp. 55--63.

\bibitem{MS}
Weiquan Fan, Zhiwei He, Xiaofen Xing, Bolun Cai, and Weirui Lu,
\newblock ``Multi-modality depression detection via multi-scale temporal
  dilated cnns,''
\newblock in {\em AVEC2019}, pp. 73--80.

\bibitem{ATransformer}
Hao Sun, Jiaqing Liu, Shurong Chai, Zhaolin Qiu, Lanfen Lin, Xinyin Huang, and
  Yenwei Chen,
\newblock ``Multi-modal adaptive fusion transformer network for the estimation
  of depression level,''
\newblock {\em Sensors}, vol. 21, no. 14, pp. 4764, 2021.

\bibitem{DepressNet}
Guramritpal~Singh Saggu, Keshav Gupta, and K.~V. Arya,
\newblock ``Depressnet: A multimodal hierarchical attention mechanism approach
  for depression detection,''
\newblock {\em IJES}, 2022.

\bibitem{CubeMLP}
Hao Sun, Hongyi Wang, Jiaqing Liu, Yen-Wei Chen, and Lanfen Lin,
\newblock ``Cubemlp: An mlp-based model for multimodal sentiment analysis and
  depression estimation,''
\newblock in {\em ACM MM}, 2022, pp. 3722--3729.

\bibitem{MultiAg}
Tao Gui, Liang Zhu, Qi~Zhang, Minlong Peng, Xu~Zhou, Keyu Ding, and Zhigang
  Chen,
\newblock ``Cooperative multimodal approach to depression detection in
  twitter,''
\newblock {\em AAAI}, vol. 33, no. 01, pp. 110--117, Jul. 2019.

\bibitem{DAIC}
Jonathan Gratch, Ron Artstein, Gale Lucas, Giota Stratou, et~al.,
\newblock ``The distress analysis interview corpus of human and computer
  interviews,''
\newblock Tech. {R}ep., University of Southern California Los Angeles, 2014.

\bibitem{CMDC}
Bochao Zou, Jiali Han, Yingxue Wang, Rui Liu, Shenghui Zhao, Lei Feng, Xiangwen
  Lyu, and Huimin Ma,
\newblock ``Semi-structural interview-based chinese multimodal depression
  corpus towards automatic preliminary screening of depressive disorders,''
\newblock {\em IEEE TAC}, 2022.

\bibitem{USE}
Daniel Cer, Yinfei Yang, Sheng-yi Kong, Nan Hua, Nicole Limtiaco, Rhomni
  St.~John, Noah Constant, Mario Guajardo-Cespedes, Steve Yuan, Chris Tar,
  Brian Strope, and Ray Kurzweil,
\newblock ``Universal sentence encoder for {E}nglish,''
\newblock in {\em EMNLP: System Demonstrations}, Brussels, Belgium, Nov. 2018,
  pp. 169--174, ACL.

\bibitem{AU2}
Ziheng Zhang, Weizhe Lin, Mingyu Liu, and Marwa Mahmoud,
\newblock ``Multimodal deep learning framework for mental disorder
  recognition,''
\newblock in {\em FG}. IEEE, 2020, pp. 344--350.

\bibitem{mlb}
Jin-Hwa Kim, Kyoung~Woon On, Woosang Lim, Jeonghee Kim, Jung-Woo Ha, and
  Byoung-Tak Zhang,
\newblock ``{Hadamard Product for Low-rank Bilinear Pooling},''
\newblock in {\em ICLR}, 2017.

\bibitem{HR}
Shi Yin, Cong Liang, Heyan Ding, and Shangfei Wang,
\newblock ``A multi-modal hierarchical recurrent neural network for depression
  detection,''
\newblock in {\em AVEC2019}, pp. 65--71.

\bibitem{resnet}
Kaiming He, Xiangyu Zhang, Shaoqing Ren, and Jian Sun,
\newblock ``Deep residual learning for image recognition,''
\newblock in {\em CVPR}, 2016, pp. 770--778.

\bibitem{eGeMaps}
George Trigeorgis, Fabien Ringeval, Raymond Brueckner, Erik Marchi, Mihalis~A
  Nicolaou, Bj{\"o}rn Schuller, and Stefanos Zafeiriou,
\newblock ``Adieu features? end-to-end speech emotion recognition using a deep
  convolutional recurrent network,''
\newblock in {\em ICASSP}. IEEE, 2016, pp. 5200--5204.

\bibitem{densenet}
Gao Huang, Zhuang Liu, Laurens Van Der~Maaten, and Kilian~Q Weinberger,
\newblock ``Densely connected convolutional networks,''
\newblock in {\em CVPR}, 2017, pp. 4700--4708.

\bibitem{CBert}
Zijun Sun, Xiaoya Li, Xiaofei Sun, Yuxian Meng, Xiang Ao, Qing He, Fei Wu, and
  Jiwei Li,
\newblock ``{C}hinese{BERT}: {C}hinese pretraining enhanced by glyph and
  {P}inyin information,''
\newblock Online, Aug. 2021, pp. 2065--2075, ACL.

\bibitem{tfo}
Gedas Bertasius, Heng Wang, and Lorenzo Torresani,
\newblock ``Is space-time attention all you need for video understanding?,''
\newblock in {\em ICML}, 2021, vol.~2, p.~4.

\bibitem{Y8}
Sami Abu-El-Haija, Nisarg Kothari, Joonseok Lee, Paul Natsev, George Toderici,
  Balakrishnan Varadarajan, and Sudheendra Vijayanarasimhan,
\newblock ``Youtube-8m: A large-scale video classification benchmark,''
\newblock {\em arXiv preprint arXiv:1609.08675}, 2016.

\end{thebibliography}

\newpage

\section{Supplymentary Material}

\subsection{Details of Feature Extraction for Text Modality}

\begin{figure}[htb]
	\centering
	\includegraphics[scale=0.14]{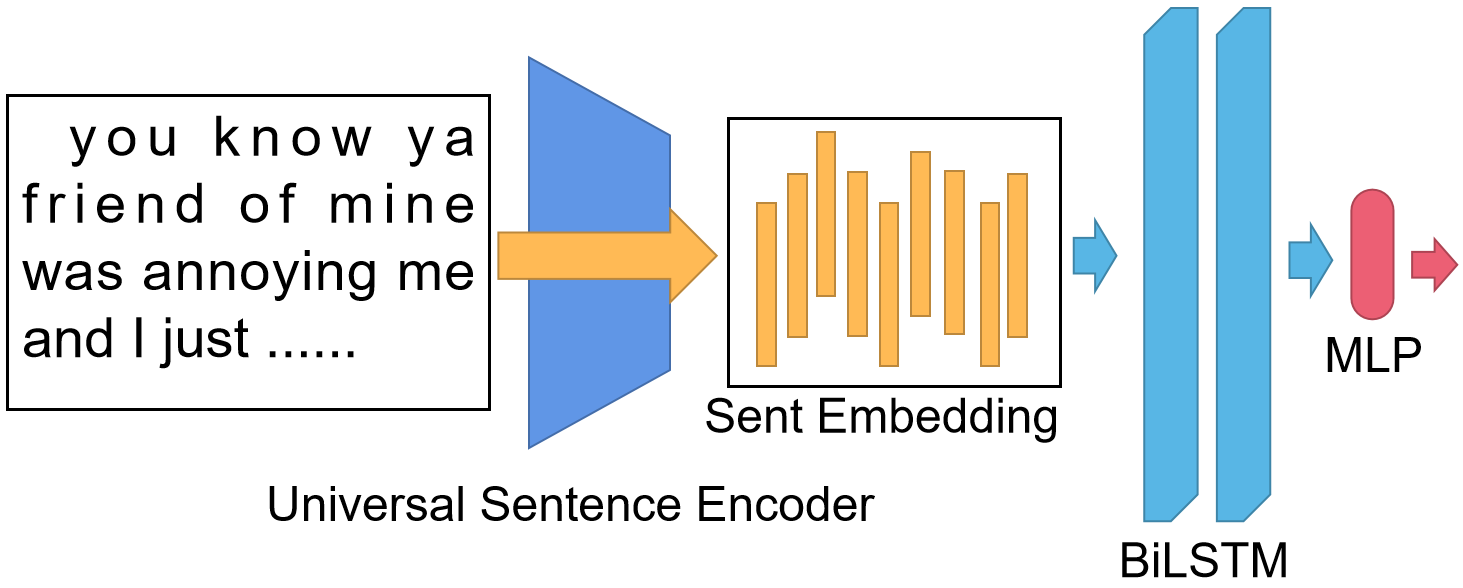}
	\caption{The adopted feature extraction pipeline for the textual interview.}
	\label{fig:text}
\end{figure}

Figure~\ref{fig:text} is the pipeline of feature extration for the patient's textual interview, where the Universal Sentence Encoder (USE)~\cite{USE} is utilized to obtain embedding for each sentence.
USE is a classic sentence to vector encoder, which obtains sentence embedding mainly by averaging the words embedding obtained from stacking transformers.
Then BiLSTM and MLP are adopted to induce a representation for the interview.

\subsection{More Experimental Details and Results}

\begin{figure*}[!t]
	\centering
    \subfigure[1st-order result]{
	\begin{minipage}{0.24\linewidth}
	\includegraphics[width=\linewidth]{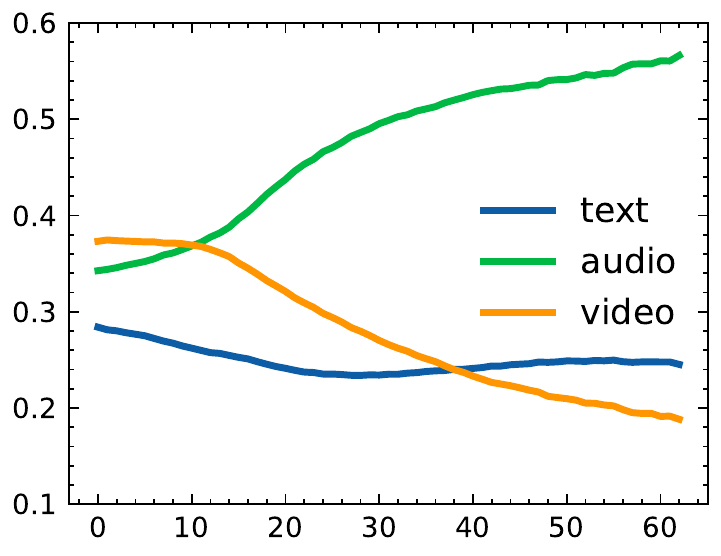}
	\label{subfig:f1}
	\end{minipage}
    }%
    \subfigure[2nd-order result]{
	\begin{minipage}{0.24\linewidth}
	\includegraphics[width=\linewidth]{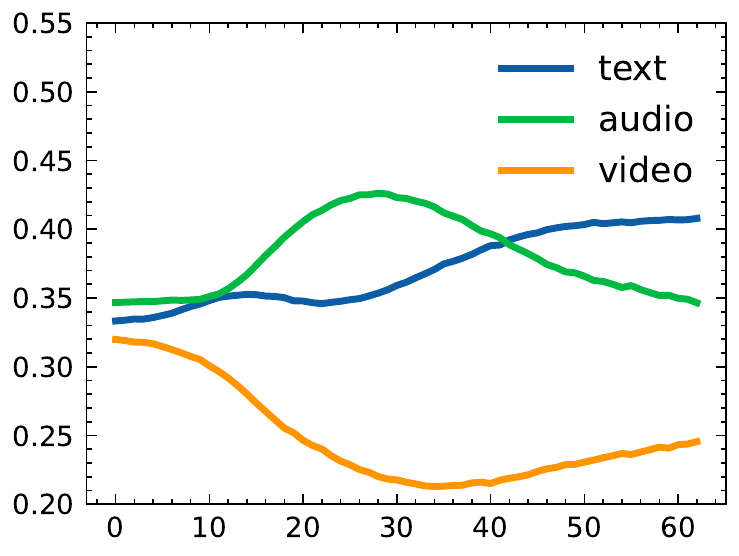}
	\label{subfig:f2}
	\end{minipage}
    }%
    \subfigure[3rd-order result]{
	\begin{minipage}{0.24\linewidth}
	\includegraphics[width=\linewidth]{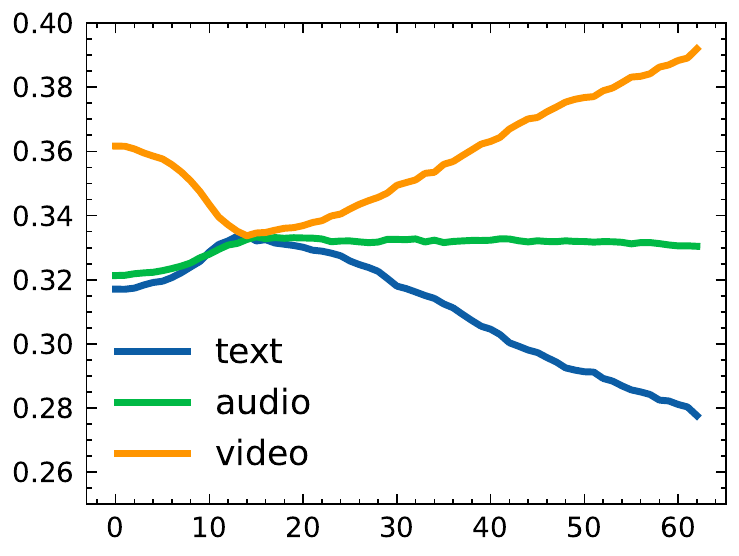}
	\label{subfig:f3}
	\end{minipage}
    }%
	\subfigure[fusion result]{
    \begin{minipage}{0.24\linewidth}
	\includegraphics[width=\linewidth]{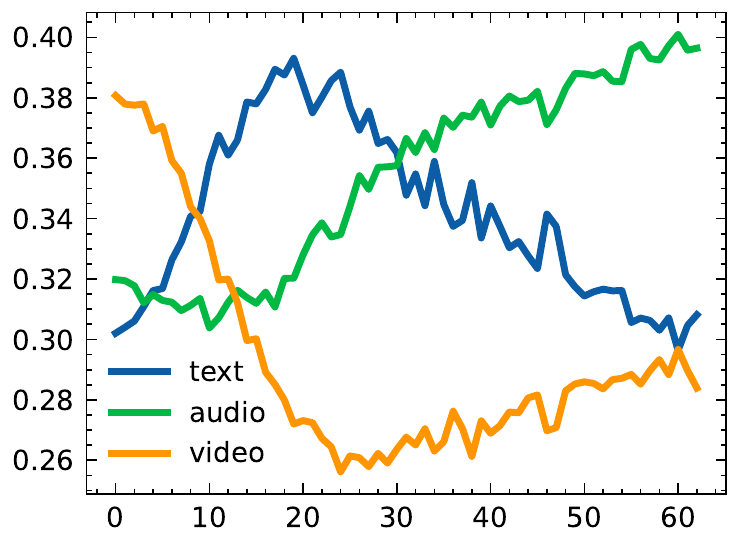}
	\label{subfig:mol}
	\end{minipage}
    }%
    \label{fig:vis}
    \caption{Visualization of the modality contributions ($\gamma^1$, $\gamma^2$, $\gamma^3$, $\gamma_{mod}$) w.r.t. the number of iterations.
    }
\end{figure*}

\subsubsection{Details of Datasets}

Our experiments are mainly conducted on two recent depression diagnosis datasets, namely, the Extended Distress Analysis Interview Corpus (E-DAIC-WOZ)~\cite{AVEC2019,DAIC} and Chinese Multimodal Compression Corpus (CMDC)~\cite{CMDC}.

The E-DAIC-WOZ is an extended version of WOZ-DAIC~\cite{DAIC}, which contain audio-visual records of patients' clinical interviews in an English environment. It is worth noting that part of the interviews in this dataset are completed through a virtual AI agent, which can effectively avoid the subjective preferences of the interviewers and is more consistent with the application scenario of automatic depression identification in the future~\cite{DAIC}. E-DAIC-WOZ provides video, audio and text modality data. For the text modality, it provides interview records. For the video modality, in order to protect the privacy of patients, the dataset only provides the extracted feature, including FAU-poses, ResNet-CNN~\cite{resnet}, etc. For the audio modality, in addition to the wav file, features extracted from MFCCs, eGeMaps~\cite{eGeMaps}, and densenet201~\cite{densenet} are also provided. In this paper, we utilize the deep-features extracted by densenet201 (audio) and ResNet-CNN (video) provided by the E-DAIC-WOZ dataset. This dataset uses the PHQ-8 (ranges in $[0, 24]$) score to evaluate the degree of depression of the patients. The larger the PHQ value, the more serious the depression of the patient.

The CMDC dataset consists of audio-visual records of patients' clinical interviews in the Chinese environment. However, unlike the continuous interview of E-DAIC-WOZ, CMDC presets a series of $12$ questions during the interview and stores the answering data for these 12 questions. For the privacy reason, the identification information in the transcribed interviews were removed. Then a ChineseBERT model~\cite{CBert} was adopted to extracted deep features. Video and audio modality data were preprocessed by some feature extractors as well, namely, pretrained TimeSFormer~\cite{tfo} for video modality and pretrained VGGish~\cite{Y8} for the audio one.
PHQ-9 score is used to evaluate the degree of depression. Considering the relatively small amount of data, the author provides a stratified 5-fold cross validation to obtain the final performance of the model.


\subsubsection{Details of Evaluation Metrics}
In the AVEC2019 DDS~\cite{AVEC2019}, CCC~(concordance correlation coefficient) and RMSE (the root mean square error) are the main evaluation metrics. To test the effectiveness of our model, we add the MAE (mean absolute error) as an extra criterion. Let $\hat{y}$ be the prediction result and $y$ the ground label, formulations of the three adopted metrics is as follows:
\begin{equation}
    \begin{split}
	& CCC = 1.0 - \frac{2Cov(y,\hat{y})}{\sigma^2_{\hat{y}} + \sigma^2_y + (\hat{y}-y)^2}, \\
	& RMSE = \sqrt{\frac{1}{N}\sum_{i=1}^N(\hat{y}-y)^2}, \\
    & MAE = \frac{1}{N} \sum_{i=1}^N |\hat{y}-y|,
	\end{split}
\end{equation}
where $\sigma^2$ indicates the variance, and $Cov$ signifies the covariance. The range of CCC is $[-1,1]$. The closer the value is to $1$, the better the fitting performance is. CCC is a metric improved from the Pearson correlation coefficient. In addition to consider the consistency/correlation, it can also measure the absolute distance difference between data.

For the CMDC dataset, in order to effectively evaluate the performance, the evaluation metrics we adopted, i.e., RMSE, MAE and Pearson correlation coefficients, are consistent with the original paper~\cite{CMDC}. The Pearson correlation coefficient can be calculated as follows:
\begin{equation}
    Pearson = \frac{Cov(\hat{y},y)}{\sigma_{\hat{y}}\sigma_y}.
\end{equation}

\subsubsection{More Ablation Studies}
In order to further evaluate the performance of MMFF and understand the fusion mechanism, we conducted ablation experiments on the E-DAIC-WOZ dataset to verify the importance of each order in MMFF. 

The sub-models are obtained by redistributing the contribution rates $\gamma$, and then removing the corresponding factors. For instance, we eliminate the 3rd-order factors in the following ways:
\begin{equation}
    \begin{split}
	& \bar{\gamma} = [\bar{\gamma}_1,\bar{\gamma}_2]^T = [\frac{\gamma_1}{\gamma_1+\gamma_2},\frac{\gamma_2}{\gamma_1+\gamma_2}]^T, \\
	& \bar{v} = \bar{\gamma}_1v^1 + \bar{\gamma}_2v^2.
	\end{split}
\end{equation}
The sub-models of excluding other orders are obtained in a similar way.

From the results of CCC in Table 3 of the main body of our paper, we can see that the general importance of order is ranked as: 1st, 3rd and 2nd. This is consistent with the results of contribution rate analysis we conduct later.

\subsubsection{More Fusion Mechanism Analysis}
In Table 4 of the main body of our paper, in addition to show the contribution of each modality in each order, we also report the general contribution of each modality in the whole MMFF process. This is calculated as:
\begin{equation}
   	\gamma_{mod} = \gamma_1 \gamma_{mod}^1 + \gamma_2 \gamma^2_{mod} + \gamma_3\gamma^3_{mod}.
\end{equation}
An visualization of the modality contribution ($\gamma^1$, $\gamma^2$, $\gamma^3$, $\gamma_{mol}$) w.r.t. the number of iterations is shown in Figure 2. It can be seen from the results that the contributions of different modalities change during the training process.
This indicates that there are complex interactions and mutual promotions between factors. For example, in a global sense (Figure \ref{subfig:mol}), the audio modality information may be promoted by the text modality.

\begin{table}[!t]
   	\centering
   	\caption{Order Contribution in MMFF, where the largest contribution is marked in bold.}
   	\label{tab:contorder}
   	\resizebox{.85\columnwidth}{!}{
   	\begin{tabular}{|l|l|l|l|}
   		   	\hline
   		   	Factor Order      & 1st Factor & 2nd Factor & 3rd Factor \\ \hline
   		   	Contribution Rate & \textbf{38.7\%}     & 29.5\%     & 31.8\%     \\ \hline
   		   	\end{tabular}
   	}
\end{table}
   	
\begin{figure}[!t]
	\centering
	\includegraphics[scale=0.7]{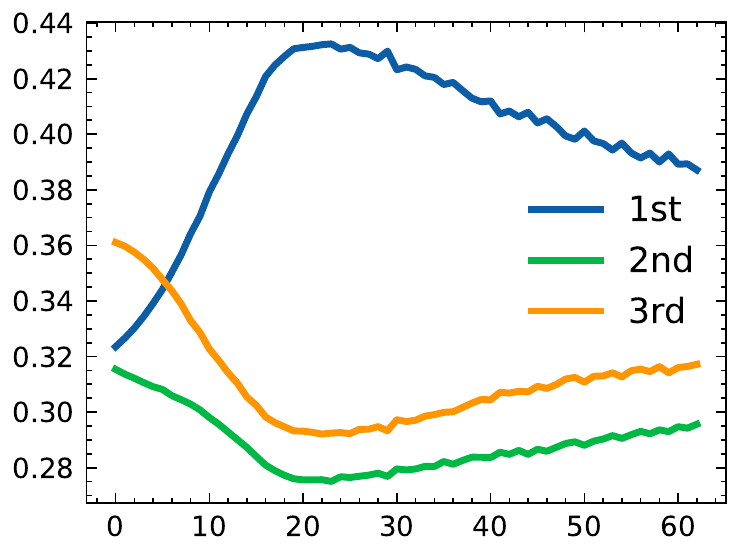}
	\caption{Order contribution w.r.t. the number of iterations in our MMFF.}
	\label{fig:order}
\end{figure} 

The contribution of each order ($\gamma$) can be analysed in the same way. The value of $\gamma $ of the fusion result is reported in Table \ref{tab:contorder} and the visualization is shown in Figure \ref{fig:order}. It can be seen that the order of factor contribution is: 1st (38.7\%), 3rd (31.8\%) and 2nd (29.5\%), which is consistent with the observations in our ablation study. As shown in Figure \ref{fig:order}, the 1st-order dominates at the early training stage. When the number of iterations increase, contributions of the 2nd and 3rd factors become larger. This may be because that the interactions between different modalities gradually increase and contributes more to the prediction.

\end{document}